\documentclass[epj,twocolumn]{webofc}
\usepackage[varg]{txfonts}   
\usepackage{amsmath}
\usepackage{bm}
\woctitle{Powders \& Grains 2017}
\begin{document}

\title{Kinetic theory of discontinuous shear thickening}
\author{\firstname{Hisao} \lastname{Hayakawa}\inst{1}\fnsep\thanks{\email{hisao@yukawa.kyoto-u.ac.jp}} \and
        \firstname{Satoshi} \lastname{Takada}\inst{2}\fnsep\thanks{\email{takada.satoshi.3s@kyoto-u.ac.jp}} 
}

\institute{Yukawa Institute for Theoretical Physics, Kyoto University, Kyoto 606-8502, Japan 
\and
           Department of Physics, Kyoto University, Kyoto University, Kyoto 606-8502, Japan 
          }

%
\abstract{%
A simple kinetic theory to exhibit a discontinuous shear thickening (DST) is proposed. The model includes a collision integral,  a drag from environment and an environmental temperature $T_{\rm ex}$ as well as a shear term. The viscosity of this model is proportional to $\dot\gamma^2$ 
for large shear rate $\dot\gamma$, while it is independent of $\dot\gamma$ for low sheae rate. 
The emergence of the DST is enhanced for lower density and lower nonzero $T_{\rm ex}$. 
}

%
\maketitle



\section{Introduction}
 The viscosity of a Newtonian fluid is independent of the shear rate $\dot\gamma$. 
This Newtonian viscosity is no longer valid, if the fluid contains some suspensions. 
 For instance, a dense suspension often behaves as a Newtonian fluid at rest or weakly stirred situation, but its resistance becomes large as if it is a solid, when the suspension is sheared with large $\dot\gamma$.
This process is known as a shear thickening which can be used in various industrial applications. 

Shear thickening attracts much attention even from physicists~\cite{Brown14,Cwalina14}.
In many of shear thickening fluids, the continuous shear thickening (CST) becomes steeper as the density increases, up to the point that the shear viscosity exhibits a discontinuous shear thickening (DST).
The suspensions exhibiting DSTs have some common features,
where
(i) the normal stress difference becomes large when the DST takes place~\cite{Cwalina14}, and (ii) it is only observed  below the jamming point~\cite{Brown09}.

To understand the DST,
Brown {\it et al.} proposed three mechanisms, hydro-clustering,  order-disorder transition and dilatancy for the shear thickening as well as their combinations~\cite{Brown14,Brown10,Waitukaitis12}. 
However, none of three mechanisms is necessary for some DSTs as in Ref.~\cite{Otsuki11}. 
Many researchers, instead, indicate
that the mutual frictions between grains play important roles in the DST mainly from simulations of suspensens and granular particles~\cite{Otsuki11,Bi11,Seto13,Mari14,Pica11,Haussinger13,Grob14,Kawasaki14}.
Nevertheless, we may ask a question whether the mutual friction is always necessary for any DST.

Some researchers~\cite{Nakanishi11,Nakanishi12,Nagahiro13,Grob16} proposed phenomenologies  for the DST by the introduction of the order parameter which exhibits a S-shape in the relationship, i. e. the flow curve, between the stress and $\dot\gamma$. 
The most successful phenomenology has been proposed by Wyart and Cates~\cite{Wyart14} in which they have introduced the fraction of lubrication film to interpolate two divergent densities for purely frictional grains and frictionless grains.
They have demontsrated that the rescaled pressure exhibit a saddle-node bifurcation at a critical density.
Then, their flow curve exhibits a S-shape above the critical density, which is an explanation of the DST.
In spite of their sucess we still do not know how to measure the fraction of lubrication film as well as the justification of their starting equation.

There are only a few microscopic theories on shear thickening. 
One of remarkable achievements is that by Santos {\it et al.}\cite{Santos98}  in which they demonstrate the existence of a CST in moderately dense hard-core gases by using the revised Enskog theory.
We do not know any systematic microscopic theory to describe DSTs, so far, as long as we know. 
Moreover, we may look for the possiblity to have a DST without the mutual friction between grains.

The purpose of this paper is to demonstrate the existence of a DST without the mutual friction between grains in terms of a simple kinetic theory.
We believe that the methodology is applicable to wide class of sheared fluids,
 though the model contains still some phenomenological assumptions.

\section{Kinetic theory: general properties}

Let us consider a collection of mono-disperse smooth spheres (the diameter $d$ and the mass $m$ without the mutual friction between spheres) influenced by the background fluid under a uniform shear flow characterized by the macroscopic velocity field $\bm{u}=(u_x,u_y,u_z)$ as
\begin{equation}
u_x=\dot\gamma y, \quad u_y=u_z=0 .
\end{equation}
Introducing the peculiar momentum $\bm{p}_i\equiv m(v_x-\dot\gamma y,v_y,v_z)$ we assume that the one-body distribution function under the simple shear satisfies
\begin{eqnarray}
& &
\left(\frac{\partial}{\partial t}+\dot\gamma
\left\{y\frac{\partial}{\partial x}-p_{y}\frac{\partial}{\partial p_{x}}
\right\}
\right)
f(\bm{r},\bm{p},t) \nonumber\\
&&=
\zeta\frac{\partial}{\partial \bm{p}} \cdot \left(
\left\{ \bm{p}+m T_{\rm ex} \frac{\partial}{\partial \bm{p}} \right\} 
 f(\bm{r},\bm{p},t) \right)
+
 J(f)
\label{boltzmann_1body} ,
\end{eqnarray} 
where we have introduced $\zeta$ and $T_{\rm ex}$ to characterize the drag from the solvent and the environmental temperature, respectively.
Here, we ignore hydrodynamic interactions between spheres.
We also assume that $\zeta$ is independent of $T_{\rm ex}$, and can be treated as a constant. 
Here, $J(f)$ is a collisional integral, which must be a functional of the two-body distribution.
Although the two-body distribution depends on the three-body distribution in general situations, 
the two-body distribution can be represented by a functional of the one-body distribution in idealistic situations such as 
dilute limit. 
Note that the position $\bm{r}$ in $f(\bm{r},\bm{p},t)$ should be interpreted as the relative position of adjacent two particles in a uniform system.
Otherwise, one can regard $f$ as the distribution function for a tracer particle.

For simplicity, 
we adopt the simple Bhatnagar-Gross-Krook (BGK) model for the collision integral for moderate dense hard spheres:~\cite{Santos98}
\begin{equation}\label{BGK}
J(f)=\frac{f_{\rm eq}-f}{\tau}
: \quad
\frac{1}{\tau}=\frac{16}{5} nd^2 g_0(\varphi) \sqrt{\frac{\pi T}{m}},
\end{equation}
where we have introduced the density $n$, the volume fraction $\varphi=\pi nd^3/6$ 
and the first peak of the radial distribution $g_0(\varphi)=(1-\varphi/2)/(1-\varphi)^3$~\cite{CS69}. 
Here the kinetic temperature $T$ introduced in Eq. \eqref{BGK} is defined by $T\equiv {\rm tr} \overleftrightarrow{\sigma}^{(K)}/(3n)$,
where $\sigma_{\alpha\beta}^{(K)}=\int d\bm{r} d\bm{p} f(\bm{r},\bm{p})p_\alpha p_\beta/m$.
Note that $T$ is the characterization of the motion of suspended particles, which differs from  $T_{\rm ex}$ for the environment.
For later convenience, we also introduce anisotropic temperatures $T_x \equiv \sigma_{xx}^{(K)}/n$.
It should be noted that the standard BGK model for moderately dense hard spheres 
should be associated with the extra terms to represent the potential contribution~\cite{Santos98}.
Thus, Eq.~\eqref{BGK} is quantitatively correct only for dilute cases. 

Now, we adopt Grad's 13 moment-like expansion~\cite{Grad49,Garzo02,Santos04},
i.\ e.\ the expansion in terms of currents such as the heat current and the stress current in addition to the conventional 5 collisional conserved quantities.
If the system is uniform, we can assume
\begin{equation}\label{Grad}
f(\bm{r},\bm{p})=f_{\rm eq}(\bm{r},\bm{p})\left\{
1+\frac{1}{2T} (\mu_{\alpha\beta}-\delta_{\alpha\beta})\tilde{\sigma}_{\alpha\beta}
\right\}
\end{equation}
where  $f_{\rm eq}=\exp[\{p^2/(2m)+U_R(r)\}/T]/{\cal Z}$ with ${\cal Z}=\int d\bm{r}\int d\bm{p}\exp[-\{p^2/(2m)+U_R(r)\}/T]$
 is the equilibrium distribution and 
\begin{equation}
\tilde{\sigma}_{\alpha\beta}\equiv \frac{p_\alpha p_\beta}{m}
- \frac{r_\alpha r_\beta}{r} U_R'(r)
\end{equation}
 is the microscopic stress. 
 Here, we adopt Einstein's rule in which duplicated Greek indices take summations over the components. 
 Note that the macroscopic stress satisfies
$\sigma_{\alpha\beta}
=\int d\bm{p}d\bm{r}f(\bm{r},\bm{p})\tilde{\sigma}_{\alpha\beta}$.
We also note that 
$U_R(r)$ is the sum of pairwise potentials, which can be replaced by the renormalized potential satisfying $U_R(r)=-T \ln g(r,\varphi)$ in the low density case, 
where $g(r,\varphi)$ is the radial distribution function~\cite{Hansen90}.
Here, we have introduced the macroscopic stress ratio
$\mu_{\alpha\beta}\equiv\sigma_{\alpha\beta}/P$ with the pressure $P\equiv {\rm tr} \overleftrightarrow{\sigma}/3$. 
This picture is consistent with the previous statistical mechanical treatments, 
where the correction term has the form $\tau \dot\gamma \tilde{\sigma}_{\alpha\beta}/T$ 
for $\alpha\ne \beta$ if $\tau\dot\gamma$ satisfies relationship $\tau\dot\gamma=\mu\equiv -\mu_{xy}=-\mu_{yx}$~\cite{Evans08,Chong10,Hayakawa13,Suzuki15}. 

Multiplying $\tilde{\sigma}_{\alpha\beta}$ on both sides of Eq.(\ref{boltzmann_1body}) with the aid of Eq.(\ref{Grad}) and integrating over $\bm{p}$ and $\bm{r}$, we obtain
\begin{equation} \label{eq:T}
\frac{\partial}{\partial t}T=\frac{1}{Z(\varphi)}
\left\{-\frac{2\dot\gamma Y(\varphi) \sigma_{xy}}{3n}+2\zeta  (T_{\rm ex}-T) \right\} , 
\end{equation}
\begin{equation}
\frac{\partial}{\partial t}\Delta T
=-\frac{2\dot\gamma Y(\varphi)\sigma_{xy}}{n Z(\varphi)}
-\left(\nu+\frac{2\zeta}{Z(\varphi)}\right) \Delta T
 \label{eq:Delta_T} , 
\end{equation}
\begin{equation}
\frac{\partial}{\partial t}\sigma_{xy}
= \dot\gamma Y(\varphi) n Z(\varphi)
\left(\frac{\Delta T}{3}-T \right) 
-\left(\nu +\frac{2\zeta}{ Z(\varphi)}\right) \sigma_{xy} ,
\label{eq:sigma_T_DT}
\end{equation}
where we have introduced
\begin{equation}\label{A(phi)}
Z(\varphi)= \frac{P}{n T}=\frac{1+\varphi+\varphi^2-\varphi^3}{(1-\varphi)^3} 
; \quad \Delta T\equiv T_x-T_y
\end{equation}
and
\begin{equation}
Y(\varphi)=\frac{2}{Z(\varphi)}-1=\frac{1-7\varphi+5\varphi^2-\varphi^3}{1+\varphi+\varphi^2-\varphi^3}.
\end{equation}
Because $Y(\varphi_c)$ becomes negative for $\varphi>\varphi_c$ with
\begin{equation}\label{phi_c}
\varphi_c=\frac{1}{3}\left(5-\frac{4}{19-3\sqrt{33}}-(19-3\sqrt{33})^{1/3} \right)
=0.160713\cdots ,
\end{equation}
our analysis is only valid for $\varphi<\varphi_c$.
It is remarkable that the normal stress is coupled with $P$ and $\sigma_{xy}$ in Eqs.~\eqref{eq:T}-\eqref{eq:sigma_T_DT}, which is consistent with Ref.~\cite{Cwalina14}. 

Thus, we obtain a closed set of equations (\ref{eq:T}) - (\ref{eq:sigma_T_DT}).
To derive Eq. \eqref{eq:sigma_T_DT} we have used
\begin{equation}\label{kinetic_potential_rel}
\sigma_{xy}^{(K)}=-\mu n T= \sigma_{xy}Z(\varphi)^{-1} 
,
\end{equation}
where we note $\sigma_{\alpha\beta}^{(K)}
=-(\mu/(m^2T)) \int d\bm{p} p_\alpha^2p_\beta^2f_{\rm eq}(\bm{p})
=-\mu n T$ and $\sigma_{xx}-\sigma_{yy}=Z(\varphi) n \Delta T$.

The damping frequency $\nu$ has been introduced in Eqs. \eqref{eq:T}-\eqref{eq:sigma_T_DT} by the relation
$\Lambda_{xy}=\int d\bm{p} d\bm{r} \tilde{\sigma}_{xy}f/\tau=\nu \sigma_{xy}$.
Note that the contribution from the diagonal term is zero for the BGK model \eqref{BGK}. 
Thanks to the assumption (\ref{Grad}), it is straightforward to show
\begin{equation}\label{nu}
\nu^*=\frac{D(\varphi)}{Z(\varphi)}\sqrt{T^*} (1+C(\varphi))
\end{equation}
for $\nu^*\equiv \nu/\zeta$ and $T^*=T/(md^2\zeta^2)$ with $D(\varphi)\equiv 48\varphi g_0(\varphi)/5\sqrt{\pi}$.
Although the precise estimation of the potential contribution $C(\varphi)$ is not easy, 
this term becomes zero if the system size is enough large and the stiffness of particles is finite.
Therefore, we omit the contribution of $C(\varphi)$ in the later discussion.

It should be noted that Eqs. \eqref{eq:T}-\eqref{eq:sigma_T_DT} cannot determine $T$, $\Delta T$ and $\sigma_{xy}$ as functions of $n$, $\dot\gamma$,  and $\nu$ because all of them are homogeneous equations.
Instead, we obtain the relations e.g. $\mu$, $\Delta T/T$ and $\dot\gamma$ as functions of $\varphi$ and $\nu$.
We also note that the motion of all particles stops, i.e. there is no dynamic steady state in the absence of $\dot\gamma$ and $T_{\rm ex}$.

In the steady state, we immediately obtain
\begin{equation}\label{DT/T}
\frac{\Delta T}{T}=\frac{6(1-\theta^{-1})}{\nu^* Z(\varphi)+2 }
\end{equation}
from Eqs.(\ref{eq:T}) and (\ref{eq:Delta_T}), 
where we have introduced $\theta\equiv T/T_{\rm ex}$. 
Substituting this into Eq.(\ref{eq:T}) we obtain
\begin{equation}\label{sigma_xy}
 \sigma_{xy}=-3\frac{1-\theta^{-1}}{\dot\gamma^* Y(\varphi)} nT.
\end{equation}
From Eqs.(\ref{eq:sigma_T_DT}), (\ref{DT/T}) and \eqref{sigma_xy} we obtain
\begin{equation}
\label{dgamma3}
\dot\gamma^*=\frac{\nu^*Z(\varphi)+2}{Y(\varphi)Z(\varphi)}
\displaystyle\sqrt{
\frac{3(\theta-1)}{\theta\nu^*Z(\varphi)+2}
}
.
\end{equation}
Then, the dimensionless viscosity is given by
\begin{equation}
\eta^*=\frac{6\varphi Y(\varphi)Z(\varphi)^2T_{\rm ex}^*\{\theta \nu^*Z(\varphi)+2\}}{\pi(\nu^* Z(\varphi)+2)^2} ,
\label{eq:eta*}
\end{equation}
where $\eta^*=\eta d/(m\zeta)$ with $\eta\equiv -\sigma_{xy}/\dot\gamma$ and $T_{\rm ex}^*=T_{\rm ex} \zeta^2/(md^2)$.
Substituting Eq.~\eqref{dgamma3} into Eq.~\eqref{sigma_xy} we obtain
\begin{equation}\label{friction_coeff}
\mu=-\frac{\sigma_{xy}}{Z(\varphi)n T}
=\frac{\sqrt{3(\theta-1)(\theta \nu^*Z(\varphi)+2)}}{\nu^*Z(\varphi)+2} 
\end{equation}
for $\theta>1$.
 Note that there is no steady solution for $\theta\le 1$, 
because the system is heated by the shear. 

{\section{Rheology at $T_{\rm ex}=0$}

Now, let us determine the expression of $\eta^*$ at $T_{\rm ex}^*=0$.
Note that $\nu^*$ can be determined by Eq. \eqref{dgamma3} as
$\sqrt{\nu^*}Z(\varphi)^{3/2}Y(\varphi)\dot\gamma^*=\sqrt{3}(\nu^* Z(\varphi)+2)$ with $\dot\gamma^*=\dot\gamma/\zeta$, whose solution is given by
$\sqrt{\nu^*}=\dot\gamma^*Y(\varphi)\sqrt{Z(\varphi)}/(2\sqrt{3})\{1\pm \sqrt{1-24/(Y(\varphi)Z(\varphi)\dot\gamma^*)^2}$.
Therefore, we obtain
\begin{equation}\label{bar_nu_1}
\nu^*
=\frac{Z(Y\dot{\gamma}^*)^{2}}{6}
\left\{
1-\frac{12}{(YZ\dot\gamma^*)^2}\pm \displaystyle\sqrt{
1-\frac{24}{(YZ\dot\gamma^*)^2}
}
 \right\} .
\end{equation}
Because $\nu^*$ must be positive, $\nu^*$ exists only if $\dot\gamma^*\ge \dot\gamma_c^*\equiv 2\sqrt{6}/(Z(\varphi)Y(\varphi))$.
Equation \eqref{bar_nu_1} indicates the absence of any dynamic steady state for $\dot\gamma^*<\dot\gamma_c^*$ because of the existence of damping terms in kinetic equation \eqref{boltzmann_1body} proportional to $\zeta$.
Equation \eqref{bar_nu_1} is also characteristics that there are two branches in $\nu^*$. 
Note that Eq. \eqref{bar_nu_1} is independent of the constitutive equation \eqref{nu}.
 
From Eqs. \eqref{nu} and \eqref{bar_nu_1} we obtain 
\begin{equation}\label{T*}
T^*=\left(\frac{5\sqrt{\pi} Z(\varphi)}{48 \varphi g_0(\varphi)
} \right)^2\nu^{*2}
\approx 
\left(\frac{5\sqrt{\pi}(Y(\varphi)Z(\varphi)\dot\gamma^*)^2 }{144\varphi g_0(\varphi) } \right)^2
\end{equation}
for $\dot\gamma^*\gg 1$.
Then, $\eta^*$ satisfies
\begin{equation}\label{eta*}
\eta^*
=
\frac{25\pi Z(\varphi)^5\nu^{*3}}{384 \varphi g_0(\varphi)^2 (\nu^*Z(\varphi)+2)^2
}
\approx 
 \frac{25\pi Z(\varphi)^4Y(\varphi)^2  \dot\gamma^{*2}}{1152\varphi g_0(\varphi)}
\end{equation}
for $\dot\gamma^*\gg 1$, where we have used Eqs.(\ref{BGK}), \eqref{nu} and \eqref{T*}.
Note that Eq.\eqref{eta*} recovers the numerical observation in Ref.~\cite{Kawasaki14} for $\dot\gamma^*\gg 1$, 
while smaller $\eta^*$ corresponding to smaller $\nu^*$ in Eq. \eqref{bar_nu_1} becomes zero for large $\dot\gamma^*$.

\begin{figure}
  \includegraphics[width=85mm]{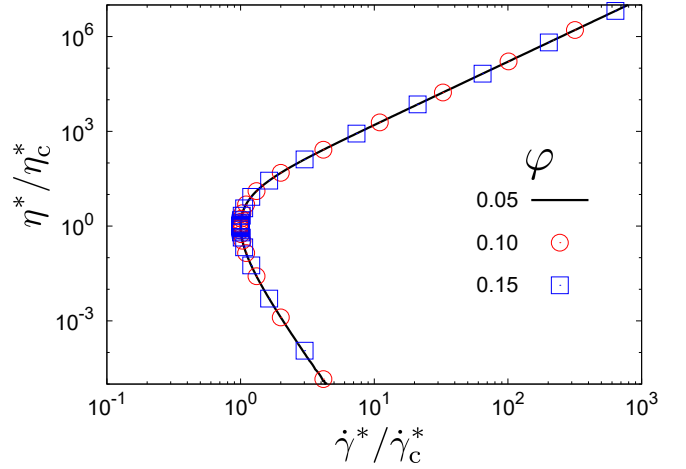}
  \caption{
  Plot of the dimensionless viscosity $\eta^*/\eta^*_c$ against the shear rate $\dot\gamma^*/\dot\gamma^*_c$ for various $\varphi$, where we adopt $T_{\rm ex}=0$.
  }
  \label{fig1}
\end{figure}

Figure \ref{fig1} is the plot of $\eta^{*}/\eta^*_c$ against $\dot\gamma^*/\dot\gamma_c^*$ for various $\varphi$, where $\eta_c^*$ is $\eta^*$ at $\dot\gamma_c^*$ based on Eqs. \eqref{dgamma3} and \eqref{eq:eta*}.
It is remarkable that the scaled flow curve is independent of $\varphi$.
This figure exhibits the existence of two branches for $\eta^*$ above $\dot\gamma_c^*$, while
 the nonexistence of $\eta^*$ below $\do\gamma_c^*$. 
The viscosity sharply increases near $\gamma_c^*$ in the upper branch corresponding to the CST but there is no DST at $T_{\rm ex}=0$.


\section{Rheology at finite $T_{\rm ex}$}   

Now, let us discuss the behavior at finite $T_{\rm ex}$.
Figure \ref{fig2} is the plots of the viscosities $\eta^*/\eta^*_c$ against $\dot\gamma^*/\dot\gamma^*_c$ for various $\varphi$. 

It is remarkable that the viscosity takes a S-shape in the flow curve at $T_{\rm ex}^* =0.1$.
The appearance of S-shape is the result of gluing of two branches or the saddle-node bifurcation, as indicated by Ref.~\cite{Wyart14}.
This S-shape is expected to appear if the Newtonian viscosity $\eta_0^*\equiv 6\varphi Z(\varphi)^2 T_{\rm ex}^*/\{\pi D(\varphi)(2+D(\varphi)\sqrt{T_{\rm ex}^*} \}$
is less than 
$\eta_c^*=25 \pi Z(\varphi)/(48 \varphi g_0(\varphi)^2)$.

More precise condition of the DST is given by $\partial \eta^*/\partial \dot\gamma^*=(\partial \eta^*/\partial \theta)/(\partial \dot\gamma^*/\partial \theta)\to \infty$.
Therefore, the critical condition $\partial \dot\gamma^*/\partial \theta=0$ can be read
\begin{equation}\label{eq:B}
8\theta^{1/2}+B^2\theta^{3/2}(1+\theta)-2B(2-7\theta+\theta^2)=0,
\end{equation}
which is a quintic equation for $\sqrt{\theta}$, where $B\equiv D(\varphi)\sqrt{T_{\rm ex}^*}$.
This S-shape in the flow cure is the requirement condition to have a DST. 

It is straightforward to analyze the linear stability of the steady solution.
The result is counter intuitive, because all steady solutions are linearly stable.
Nevertheless, when we control $\dot\gamma$, the viscosity discontinuously changes, if the flow curve has the S-shape.  
The amount of discontinuity in the flow curve is finite as shown in Fig. \ref{fig2}.

\begin{figure}
  \includegraphics[width=85mm]{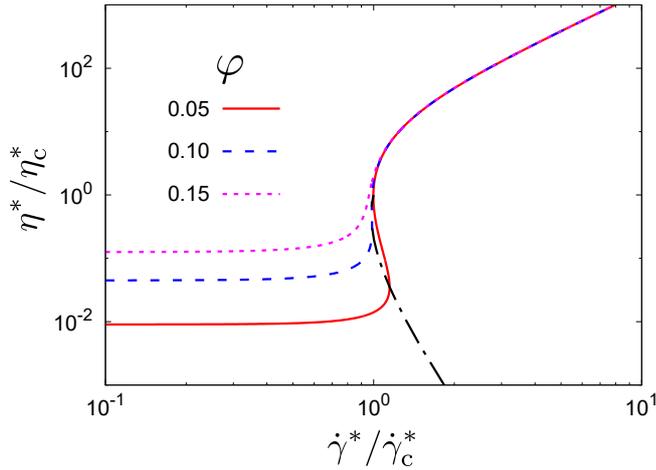}
  \caption{
  Plot of  $\eta^*/\eta_c^*$ against $\dot\gamma^*/\dot\gamma_c^*$ at $T_{\rm ex}^*=0.1$ for various $\varphi$. 
  The back dot-dashed line satisfies $\partial \dot\gamma^*/\partial \eta^*=0$.
  }
  \label{fig2}
\end{figure}

\section{Discussion and Conclusion} 

Now, let us discuss our results.
We should note that the DST in our model takes place for $\varphi<\varphi_c(T_{\rm ex}^*)$.
This result is contrast to experimental observations, though our result is obtained from
the fact that the Newtonian viscosity $\eta_0^*$ becomes lower if $\varphi$ decreases.
After the completion of this work, we have confirmed that the Boltzmann equation instead of BGK model in the dilute limit exhibits a DST which perfectly agrees with the simulaion result~\cite{arXiv1611} and the DST disappears as the density increases from the analysis of both the Enskog equation and the simulation~\cite{Enskog17}.  

One of key assumptions in our analysis is Eq.~\eqref{Grad}.
Fortunately, Eq. \eqref{Grad} seems to be consistent with the Green-Kubo formula in nonequilibrium statistical mechanics~\cite{Evans08,Chong10,Hayakawa13,Suzuki15}.

We have assumed that the system is uniform to simplify the argument.
This assumption cannot be used if the phase separation takes place.

In conclusion, we demonstrate that our simple kinetic model which has a constant damping $\zeta$ and an external temperature $T_{\rm ex}$ exhibits the DST.
The emergence of the DST is the result of a saddle-node bifurcation by the gluing between Newtonian branch and the branch at $T_{\rm ex}=0$. 
We also note that the normal stress difference plays an important role in the DST,
which is consistent with Ref.~\cite{Cwalina14}.
The proposed mechanism is new, because the DST is unrelated to neither mutual friction of grains, the order-disorder transition, boundary effects nor hydro-clustering mechanism. 

{\bf Acknowledgment}:
We thank K. Suzuki, T. Kawasaki, M. Otsuki, K. Saitoh and R. Seto for his useful comments. This work is partially supported by the Grant-in-Aid of MEXT for Scientific Research (Grant No. 16H04025).

\end{document}